\documentclass[12pt]{iopart}

\usepackage{graphicx}
\usepackage[colorlinks=true,citecolor=blue]{hyperref}
\usepackage{ulem}
\usepackage{txfonts}
\usepackage{iopams}

\newcommand{\ket}[1]{\ensuremath{|{#1}\rangle}}

\begin{document}

\title{Stimulated Neutrino Transformation with Sinusoidal Density Profiles}

\author{J P Kneller, G C McLaughlin and K M Patton}
\address{Department of Physics, North Carolina State University,Raleigh,North Carolina 27695-8202, USA}

\ead{\mailto{jim\_kneller@ncsu.edu}, \mailto{gail\_mclaughlin@ncsu.edu} \mailto{kmpatton@ncsu.edu}}

\begin{abstract}
Large amplitude oscillations between the states of a quantum system can be
stimulated by sinusoidal external potentials with frequencies that are similar
to the energy level splitting of the states or a fraction thereof. Situations when the applied frequency is equal to an integer fraction of the
energy level splittings are known as parametric resonances. We investigate this effect for neutrinos both analytically and numerically 
for the case of arbitrary numbers of neutrino flavors. We look for environments where the effect may be observed and find that supernova are the one realistic possibility 
due to the necessity of both large densities and large amplitude fluctuations. The comparison of numerical and analytic results of neutrino propagation through a model supernova reveals 
it is possible to predict the locations and strengths of the stimulated transitions that occur.
\end{abstract}

\maketitle

%
%
%

\section{Introduction}

Neutrino flavor transformation is a complex phenomenon: the general coherent neutrino state is a linear combination of several states with different energy and 
the oscillations between these states are affected by the presence of matter and may exhibit many-body effects, see the review by Duan \& Kneller \cite{Duan:2009cd}, 
if the neutrino density is sufficiently high. 
Several methods for finding the solution through a general varying density profile exist, e.g.\ \cite{1996PhRvD..54.6323B,2001JPhG...27.2405F,2001EPJC...20..507O,2004JMP....45.4053B},  
but if we focus upon the specific case of a periodic profile we find a neutrino flavor transformation phenomenon similar to stimulated transitions between 
atomic or molecular states driven by an external perturbing potential \cite{Ermilova,1987PhLB..185..417S,Akhmedov,1989PhLB..226..341K,PhysRevD.43.2484,1996PhRvD..54.3941B,2009PhLB..675...69K}.
It was found, for example, by Akhmedov \cite{1999NuPhB.538...25A} that large amplitude transitions - called parametric resonances - occur between the neutrino states when the frequency of the perturbation, $k_{\star}$, matches an integer fraction $n$ of the frequency associated with one of the energy splittings $\delta k_{ab}$, i.e.\ $n\,k_{\star} = |\delta k_{ab}|$. 
This result is intriguing because any density profile can be decomposed into a `smooth' profile of some kind and the `fluctuations' can be represented by a Fourier series. One may expect that 
some terms in the series may have wavelengths which are capable of stimulating transitions.
Specific discussions of the effect of `fluctuations' in the Earth are found in Akhmedov \cite{1999NuPhB.538...25A,2001PAN....64..787A} and Jacobsson \emph{et al.} \cite{2002PhLB..532..259J}, for the Sun the reader is referred to Loreti \emph{et al.} \cite{Loreti:1994ry} and  Haxton and Zhang \cite{PhysRevD.43.2484}, and for supernovae there are many papers in the literature to consult \cite{Loreti:1995ae,Kneller:2010ky,Kneller:2010sc,Friedland:2006ta,2003PhRvD..68c3005F}. 

In this paper we revisit the phenomenon of stimulated transition and derive new analytic results valid both on and off the 
parametric resonance, at all ambient densities and for arbitrary number of neutrino flavors. 
After obtaining the new expressions and testing them against numerical solutions, we estimate the possibilities for observing stimulated transitions in a terrestrial experiment, in the Sun and in supernovae. 
After finding that stimulated transitions are likely only to occur in supernovae we proceed to a supernova test problem using an analytic supernova density profile 
upon which we superpose a perturbing sinusoidal fluctuation. We predict the places where the transitions will occur and their strength and then verify our predictions with a numerical 
three neutrino oscillation calculation.

%
%
%

\section{Stimulated Neutrino Transformation}

We are interested in the probability that some
initial neutrino state $\ket{\nu(r)}$ at $r$ is later detected as the
state $\ket{\nu(r')}$ at $r'$. These probabilities are computed from the $S$-matrix 
which relates the initial and final states via $\ket{\nu(r')} =
S(r',r)\,\ket{\nu(r)}$. The probabilities we calculate depend upon the basis and to avoid the 
intrinsic oscillatory behaviour in the flavor basis we prefer to use the instantaneous eigenstates of the Hamiltonian $H$, 
known as the matter basis. The two bases are related by the unitary transformation matrix $U$. For two flavors/mass states the 
matrix $U$ is parametrised by a single mixing angle, $\theta$; for three flavors/mass states the 
matrix $U$ is parametrised by three mixing angles, $\theta_{12}$, $\theta_{13}$
and $\theta_{23}$, a CP phase and two, irrelevant, Majorana phases. 
The differential equation for the neutrino $S$-matrix \cite{Kneller:2005hf,Kneller:2009vd} is simply
\begin{equation}
\imath \frac{dS}{dr} = H\,S \label{eq:dSdr}
\end{equation}
and in the flavor basis the Hamiltonian is $H^{(f)} = U_{0}\,K^{(m)}_{0}\,U_{0}^{\dagger} + V^{(f)}$ where $K^{(m)}_{0}$ is a diagonal matrix in the vacuum, $U_{0}$ the vacuum
mixing matrix, and $V^{(f)}(r)$ some `potential' that we allow to be position dependent. Our goal is 
to solve \eref{eq:dSdr} for $S$ given some potential $V(r)$. 
The problem we have in mind is the case where $V(r)$ possesses some sort of `smooth' component we denote by $\breve{V}$ and a perturbation $\delta V$. 
Thus the flavor basis Hamiltonian is split into three terms, $H^{(f)} =
U_{0}\,K^{(m)}_{0}\,U_{0}^{\dagger} + \breve{V}^{(f)} + \delta V^{(f)}$, 
which we group as $H^{(f)} = \breve{H}^{(f)} + \delta V^{(f)}$.
The matter states are defined as those which diagonalise $H^{(f)}$ i.e.\ if
$K^{(m)}$ is the diagonal matrix of eigenvalues of $H$ then $U$ - the matter mixing matrix - is $H^{(f)} = U\,K^{(m)}\,U^{\dagger}$. 
But we also define another basis - the unperturbed matter basis. If $\breve{K}^{(\breve{m})}$ is the diagonal matrix of unperturbed 
eigenvalues, $\breve{k}_{1}, \breve{k}_{2},\ldots$, of $\breve{H}$ then we we can introduce an unpeturbed mixing matrix $\breve{U}$ defined by 
$\breve{H}^{(f)} = \breve{U}\,\breve{K}^{(\breve{m})}\,\breve{U}^{\dagger}$. In this unpeturbed matter basis
\begin{equation}
H^{(\breve{m})} = \breve{K}^{(\breve{m})} - \imath
\breve{U}^{\dagger}\,\frac{d\breve{U}}{dr} +
\breve{U}^{\dagger}\delta V^{(f)}\breve{U} 
\end{equation} 
We now write the $S$-matrix for the unpeturbed matter basis as the product $S^{(\breve{m})} = \breve{S}\,A$ where
$\breve{S}$ is defined to be the solution of 
\begin{equation}
\imath \frac{d\breve{S}}{dr} = \left[ \breve{K}^{(\breve{m})} - \imath
\breve{U}^{\dagger}\,\frac{d\breve{U}}{dr} \right] \,\breve{S}.
\end{equation}
If we know the solution to the unperturbed problem, $\breve{S}$, we can solve for the effect of the 
perturbation by finding the solution to the differential equation for $A$: 
\begin{equation}
\imath \frac{dA}{dr} = \breve{S}^{\dagger}\,\breve{U}^{\dagger}\delta V^{(f)}\breve{U}\,\breve{S} \,A. \label{dAdr}
\end{equation}
In general the term $\breve{U}^{\dagger}\delta V^{(f)}\breve{U}$ which appears in this equation possesses both
diagonal and off-diagonal elements. The diagonal elements are easily removed by 
writing the matrix $A$ as $A=W\,B$ where $W=\exp(-\imath\Xi)$ and $\Xi$ a diagonal matrix
$\Xi=diag(\xi_{1},\xi_{2},\ldots)$. Substitution into \eref{dAdr} gives a differential equation for $B$
\begin{equation}
\imath \frac{dB}{dr} = W^{\dagger}\left[\breve{S}^{\dagger}\breve{U}^{\dagger}\delta V^{(f)}\breve{U}\,\breve{S} -\frac{d\Xi}{dr}\right]\,W\,B
\end{equation}
and $\Xi$ is chosen so that $d\Xi/dr$ removes the diagonal 
elements of $\breve{S}^{\dagger}\breve{U}^{\dagger}\delta V^{(f)}\breve{U}\,\breve{S}$.
Once $\Xi$ has been found, determining transition probabilities involves solving $dB/dr$.

%
%
%
\section{Constant Potentials with Sinusoidal Perturbations}
\label{sec:constantpotential}

We now consider the specific case of a matter potential, so that the only non-zero entry
of $\breve{V}^{(f)}$ is $\breve{V}_{ee}$ and $\breve{V}_{ee}(r)$ is $\breve{V}_{ee}(r) = \sqrt{2}\,G_{F}\,\breve{n}_{e}(r)$ where $\breve{n}_{e}$ is the `smooth' electron density. 
With this form for the potential the elements of $\breve{U}^{\dagger}\breve{V}^{(f)}\breve{U} = \breve{V}^{(\breve{m})}$ are
$\breve{V}^{(\breve{m})}_{ij} = \breve{U}_{ei}^{\star}\,\breve{U}_{ej}\,V_{ee}$. Next we Taylor expand $\breve{V}_{ee}(r)$ as $\breve{V}_{ee}(r) = V_{\star} + \ldots$ and retain only the first term of the expansion. For $\breve{V}_{ee}(r)=V_{\star}$ the S-matrix describing the unperturbed problem, $\breve{S}$, is diagonal because the eigenvalues of $\breve{H}$ and
thus the mixing matrix $\breve{U}$ are fixed so $\breve{S}$ is simply $\breve{S} = \exp\left(-\imath\breve{K}^{(\breve{m})}\,r\right)$. 
The perturbation we consider is the case of a single sinusoidal fluctuation of wavenumber $k_{\star}$, amplitude $C_{\star}$ and phase shift $\eta$ i.e.\ $\delta V_{ee}(r) = C_{\star}\,V_{\star} \sin\left(k_{\star}\,r+\eta\right)$. With this perturbing potential we solve for the quantities $\xi_{i}$:
\begin{equation}
\xi_{i} =
\frac{C_{\star}\,V_{\star}\,|\breve{U}_{ei}|^{2}}{k_{\star}}\left[\cos\eta -
\cos\left(k_{\star}\,r+\eta\right)\right]. 
\end{equation}
Everything we require to start solving the equation for $dB/dr$ has now been defined.
The differential equation for the matrix $B$ is 
\begin{eqnarray}\label{eq:idBdr}
\fl \imath \frac{dB}{dr} = C_{\star}\,V_{\star}\,\sin\left(k_{\star}\,r+\eta\right) & & \nonumber \\  
   \times \,\left( \begin{array}{cccc}
     0 & \breve{U}_{e1}^{\star}\breve{U}_{e2} e^{\imath\left(\delta\breve{k}_{12} r+\delta\xi_{12}\right)} & \breve{U}_{e1}^{\star}\breve{U}_{e3} e^{\imath\left(\delta\breve{k}_{13} r+\delta\xi_{13}\right)} & \ldots\\
     \breve{U}_{e2}^{\star}\breve{U}_{e1}e^{-\imath\left(\delta\breve{k}_{12} r+\delta\xi_{12}\right)} & 0 & \breve{U}_{e2}^{\star}\breve{U}_{e3} e^{\imath\left(\delta\breve{k}_{23} r+\delta\xi_{23}\right)} & \ldots \\
     \breve{U}_{e3}^{\star}\breve{U}_{e1} e^{-\imath\left(\delta\breve{k}_{13} r+\delta\xi_{13}\right)} & \breve{U}_{e3}^{\star}\breve{U}_{e2} e^{-\imath\left(\delta\breve{k}_{23} r+\delta\xi_{23}\right)} & 0 & \ldots \\
     \vdots & \vdots & \vdots & \ddots
     \end{array} \right)\,B & &  
\end{eqnarray}
where $\delta\breve{k}_{ij} = \breve{k}_{i} -\breve{k}_{j}$ and similarly $\delta\xi_{ij}=\xi_{i}-\xi_{j}$. 
Our first step in solving this equation is to introduce the quantities $z_{ij}$ and $\kappa_{ij,n}$, defined to be  
\begin{eqnarray}
z_{ij} = \frac{C_{\star}\,V_{\star}}{k_{\star}} \left(|\breve{U}_{ei}|^{2}-|\breve{U}_{ei}|^{2}\right), \\
\kappa_{ij,n} = (-\imath)^{n-1}\,\frac{n\,C_{\star}\,V_{\star}}{z_{ij}}\,J_{n}(z_{ij})
\breve{U}_{ei}^{\star}\breve{U}_{ej} \,e^{\imath\left(n\, \eta +z_{ij} \cos\eta \right)} 
\end{eqnarray}
respectively where $n$ is an integer and $J_{n}$ is the Bessel J function, 
and our second is to use the Jacobi-Anger expansion for $\exp\left(\imath\delta \xi_{ij}\right)$ 
\begin{equation}\label{eq:JAexpansion}
\exp\left(\imath\delta \xi_{ij}\right) = \exp\left(\imath\, z_{ij}\cos\eta\right) 
\sum_{n=-\infty}^{\infty} (-\imath)^{n} J_{n}\left(z_{ij}\right)\exp\left[\imath\,n\left(k_{\star}\,r+\eta\right)\right]
\end{equation}
With this expansion and two definitions we find \eref{eq:idBdr} becomes 
\begin{equation} \label{eq:dBdr:offres}
\fl \imath \frac{dB}{dr} = \imath \sum_{n=-\infty}^{\infty} 
\left( \begin{array}{cccc}
    0 & -\kappa_{12,n} e^{\imath\left(\delta\breve{k}_{12} + n\,k_{\star}\right) r} & -\kappa_{13,n} e^{\imath\left(\delta\breve{k}_{13} + n\,k_{\star}\right) r} & \ldots \\
    \kappa^{\star}_{12,n} e^{-\imath\left(\delta\breve{k}_{12} + n\,k_{\star}\right) r} & 0 & -\kappa_{23,n} e^{\imath\left(\delta\breve{k}_{23} + n\,k_{\star}\right) r} & \ldots \\
    \kappa^{\star}_{13,n} e^{-\imath\left(\delta\breve{k}_{13} + n\,k_{\star}\right) r} & \kappa^{\star}_{23,n} e^{-\imath\left(\delta\breve{k}_{23} + n\,k_{\star}\right) r} & 0 & \ldots \\
    \vdots & \vdots & \vdots & \ddots
    \end{array}\right)\,B  
\end{equation} 
which is a greatly simplified equation. 


We now adopt the Rotating Wave Approximation (RWA) and for each element of the matrix drop all terms in the series appearing in \eref{eq:dBdr:offres} 
except the one closest to parametric resonance i.e.\ we find the integer $n_{ij}$ which satisfies $\delta\breve{k}_{ij} + n_{ij}\,k_{\star} \approx 0$ 
for each eigenvalue splitting $\delta\breve{k}_{ij}$.
With this approximation we find that \eref{eq:dBdr:offres} falls into a
class of matrix Schrodinger equations with known solutions. First we introduce the RWA Hamiltonian $H^{(B)}(r)$
so that our differential equation for $B$ is  
\begin{equation} \label{eq:dBdr:offresRWA} 
\imath \frac{dB}{dr}  = H^{(B)} B,
\end{equation} 
where
\begin{equation}
\fl  
H^{(B)}(r) = \left( \begin{array}{cccc}
    0 & -\imath \kappa_{12,n_{12}} e^{\imath\left[\delta\breve{k}_{12} + n_{12}\,k_{\star}\right] r} & -\imath \kappa_{13,n_{13}} e^{\imath\left[\delta\breve{k}_{13} + n_{13}\,k_{\star}\right] r} & \ldots \\
    \imath \kappa^{\star}_{12,n_{12}} e^{-\imath\left[\delta\breve{k}_{12} + n_{12}\,k_{\star}\right] r} & 0 & -\imath \kappa_{23,n_{23}} e^{\imath\left[\delta\breve{k}_{23} + n_{23}\,k_{\star}\right] r} & \ldots \\
    \imath \kappa^{\star}_{13,n_{13}} e^{-\imath\left[\delta\breve{k}_{13} + n_{13}\,k_{\star}\right] r} & \imath \kappa^{\star}_{23,n_{23}} e^{-\imath\left[\delta\breve{k}_{23} + n_{23}\,k_{\star}\right] r} & 0 & \ldots \\
    \vdots & \vdots & \vdots & \ddots
    \end{array}\right)
\label{eq:hb}
\end{equation}
The value of $n_{ij}$ which fulfills this condition will, in general, be different for each pair $ij$ but they are not all independent: 
the requirement that $\delta\breve{k}_{ab} + \delta\breve{k}_{bc} = \delta\breve{k}_{ac}$ implies 
$n_{ab} + n_{bc} = n_{ac}$. 
This identity allows us to factorise $H^{(B)}(r)$ into the form $H^{(B)}(r) = \Upsilon(r)\,H_{0}\,\Upsilon^{\dagger}(r)$ where the matrix $H_{0}$ is a constant, i.e. it
contains $\kappa_{ij,n_{ij}}$ (and its complex conjugate) but has no dependence on $r$. 
The matrix $\Upsilon$ is of the form $\Upsilon(r)=\exp(\imath\,Q\,r)$, where $Q$ is also a constant matrix but depending on $\delta\breve{k}_{ij}$ and $n_{ij}\,k_{\star}$.
Explicitly we can write 
\begin{equation}
H_{0} = \left( \begin{array}{cccc}
    0 & -\imath \kappa_{12,n_{12}} & -\imath \kappa_{13,n_{13}} & \ldots \\
    \imath \kappa^{\star}_{12,n_{12}} & 0 & -\imath \kappa_{23,n_{23}} & \ldots \\
    \imath \kappa^{\star}_{13,n_{13}} & \imath \kappa^{\star}_{23,n_{23}} & 0 & \ldots \\
    \vdots & \vdots & \vdots & \ddots
    \end{array}\right)
\end{equation}
but the matrix $Q$ is not unique. One possibility is 
\begin{equation}
Q = \left( \begin{array}{cccc}
    \breve{k}_{1} + n_{1}\,k_{\star} & 0 & 0 & \ldots \\
    0 & \breve{k}_{2} + n_{2}\,k_{\star} & 0 & \ldots \\
    0 & 0 & \breve{k}_{3} + n_{3}\,k_{\star} & \ldots \\
    \vdots & \vdots & \vdots & \ddots
    \end{array}\right)
\end{equation}
where $n_{i}$ are integers chosen so that $n_{i} - n_{j} = n_{ij}$. 
Despite this ambiguity in $Q$, the decomposition means \eref{eq:dBdr:offresRWA} can be written as   
\begin{equation}
i\Upsilon^{\dagger}\frac{dB}{dr} =  H_{0}\Upsilon^{\dagger}\,B
\end{equation}
The form of this equation suggests that instead of solving for $B$ directly we should solve for the combination $\Omega=\Upsilon^{\dagger} B$. The differential equation for $\Omega$ is found to be 
\begin{equation}
i\frac{d\Omega}{dr} = \left(H_{0} +Q \right)\,\Omega = H^{(\Omega)}\,\Omega.
\end{equation}
Since the matrix $H^{(\Omega)}$ is independent of $r$, $\Omega$ has the simple solution of $\Omega(r) = \exp(-\imath H^{(\Omega)} r)\,\Omega(0)$ which can be used to derive the solution for $B$ as  
\begin{equation}
B(r) = \Upsilon(r)\,\exp(-\imath H^{(\Omega)} r)\,\Upsilon^{\dagger}(0) B(0). \label{eq:soln for B}
\end{equation}
For the particular case of two flavors we can write out the solution succinctly after dropping 
the subscripts on $n_{12}$ and $\kappa_{12,n_{12}}$, and introducing $2k_{n} = \delta\breve{k}_{12} + n\,k_{\star}$ and $q_{n}^{2} = k_{n}^{2}+ \kappa_{n}^{2}$. Our expression for $B(r)$ for two flavors is that 
\begin{equation}
B = \left( \begin{array}{cc}
 e^{\imath k_{n} x} \left[ \cos(q_{n} x) -\imath\frac{k_{n}}{q_{n}}\sin(q_{n} x)
\right] & -e^{\imath k_{n} x}\,\frac{\kappa_{n}}{q_{n}}\sin(q_{n} x) \\
e^{-\imath k_{n} x}\frac{\kappa_{n}^{\star}}{q_{n}}\sin(q_{n} x) & e^{-\imath
k_{n} x} \left[ \cos(q_{n} x) +\imath\frac{k_{n}}{q_{n}}\sin(q_{n} x) \right] 
 \end{array} \right)
\end{equation}
Because both $\breve{S}$ and $W$ are diagonal matrices, from this solution for $B$ we can read off
the transition probability between the matter states 1 and 2 as 
\begin{equation}
P_{12} = |B_{12}|^{2}= \frac{\kappa_{n}^2}{q_{n}^2}\,\sin^{2}(q_{n} r) \equiv A_{12}\sin^{2}(q_{n} r). \label{eq:P_12}
\end{equation}
The reduced transition wavelength $\lambdabar_{n}$ is simply $\lambdabar_{n}=1/q_{n}$ and at a parametric resonance $\lambdabar_{n} = 1/|\kappa_{n}|$.
We also observe that the range of $k_{\star}$ such that the amplitude $A_{12}$ of the oscillation is greater
than $1/2$ is $\Delta k_{\star} = 4|\kappa_{n}|/n$ and for small $z_{12}$ the Bessel function can be approximated as $J_{n}(z_{12}) \propto z_{12}^{n}$ so
we see that this width of the resonance is proportional to
$C_{\star}^{n}$ when the parameter $z_{12}$ is small. In summary then, we expect to see in $P_{12}(k_{\star})$ a series of peaks at the undertones of
$|\delta\breve{k}_{12}|$ with widths that decrease geometrically. 
Our result for $P_{12}(r)$ has a similar form to equation (2) in Ermilova \etal \cite{Ermilova} but is applicable over a wider range of densities.  
It matches the equation derived by Kmetic \& Meath \cite{1985PhLA..108..340K} for the case of irradiated two-level molecules with permanent dipoles.


\subsection{Comparison with numerical results}

\begin{figure}
\includegraphics[clip,width=\linewidth]{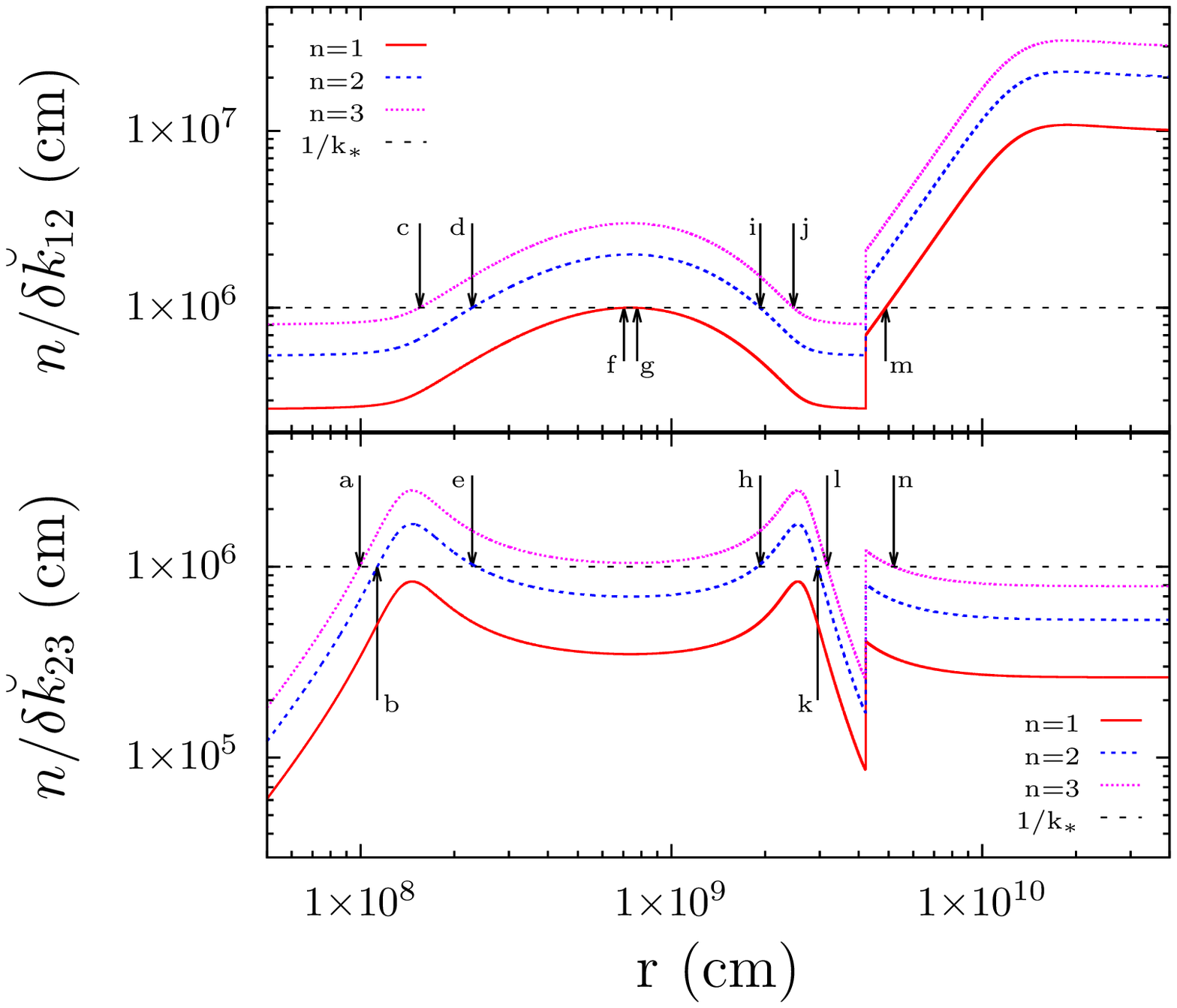}
\caption{A comparison between a numerical result (red curve) and the analytic solution (blue circles) for a particular test problem. 
The neutrino energy is $20\;{\rm MeV}$, the vacuum mixing angle $\theta=0.573^{\circ}$ and the mass splitting $\delta m^{2} = 3\times 10^{-3}\;{\rm eV^2}$. 
The density $V_{\star}$ is set to one half the Mikheev-Smirnov-Wolfenstein (MSW) \cite{Wolfenstein1977,M&S1986} density for this energy and mixing parameters and $C_{\star}=0.1$. The wavelength 
of the sinusoidal component of the density profile is set at $\lambda_{\star} = 5.27\;{\rm km}$.} \label{fig:compare}
\end{figure}
To determine whether our analytic results are useful we compare their predictions with the numerical results for a set of test problems. 
The test problem we select is the case of a neutrino with an energy of $20\;{\rm MeV}$, a vacuum mixing angle of $\theta=0.573^{\circ}$ and 
the mass splitting at $\delta m^{2} = 3\times 10^{-3}\;{\rm eV^2}$. The density $V_{\star}$ is set to one half the MSW density for this energy and mixing parameters 
and $C_{\star}=0.1$. The wavelength of the sinusoid is set to $\lambda_{\star}=5.27\;{\rm km}$ which was selected so as to be close to a parametric resonance $n=1$.

Our code solves equation (\ref{eq:dSdr}) with the potential $V(r) = V_{\star} \,\left\{ 1+ C_{\star}\,\sin\left(k_{\star}\,r+\eta\right) \right\}$ for the 
$S$-matrix using the adiabatic basis introduced in Kneller \& McLaughlin \cite{Kneller:2005hf} for the two-flavor MSW problem and then later extended to arbitrary numbers of neutrino flavors and arbitrary Hamiltonians \cite{2009PhRvD..80e3002K,2012JPhG...39c5201G}. The $S$-matrix is parametrised so as to ensure unitarity and the differential equations for its eight parameters are solved using a simple Runge-Kutta adaptive step size integrator. 
   
The comparison between the analytic and numerical results for this test problem is shown in \fref{fig:compare} and we observe that they are in very good agreement: both the wavelength and amplitude are correctly predicted. Note that the transition wavelength ($\lambda_{1} \sim 1.3 \times 10^{4}\;{\rm km}$) is much longer than the wavelength of the sinusoid ($\lambda_{\star}=5.27\;{\rm km}$) which, in this case, is also close to the wavelength we associate with the splitting between the eigenstates, which is $5.26\;{\rm km}$.    
\begin{figure}[t]
\includegraphics[clip,width=\linewidth]{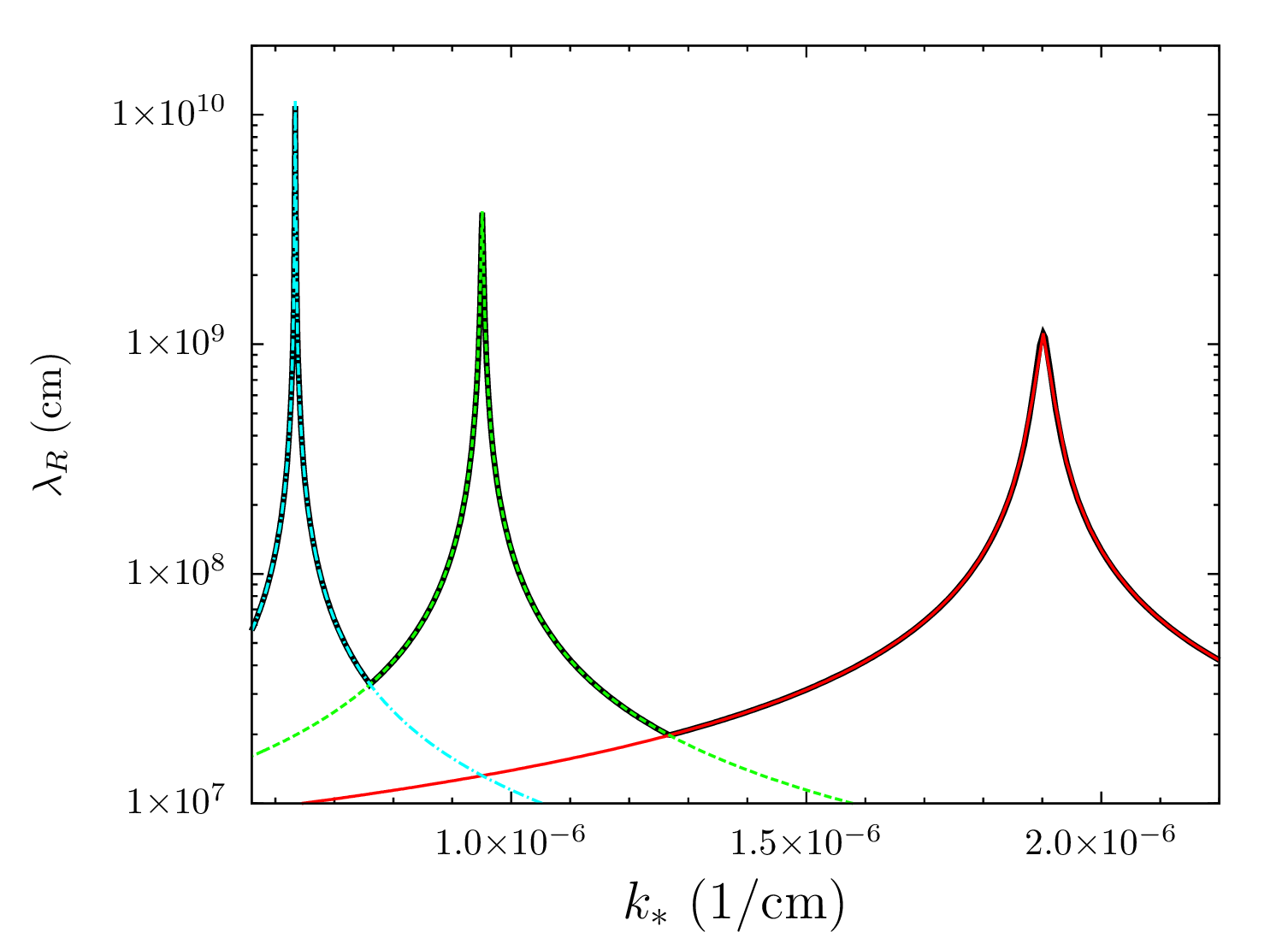}
\caption{A comparison between the Rotating Wave Approximation prediction of the oscillation wavelength for resonances $n\in\{1,2,3\}$ ($n=1$ is the rightmost peak) and the wavelength derived from the numerical solution (solid black) as a function of the applied frequency of our test problem.} \label{fig:compare_wavelength}
\end{figure}
\begin{figure}[t]
\includegraphics[clip,width=\linewidth]{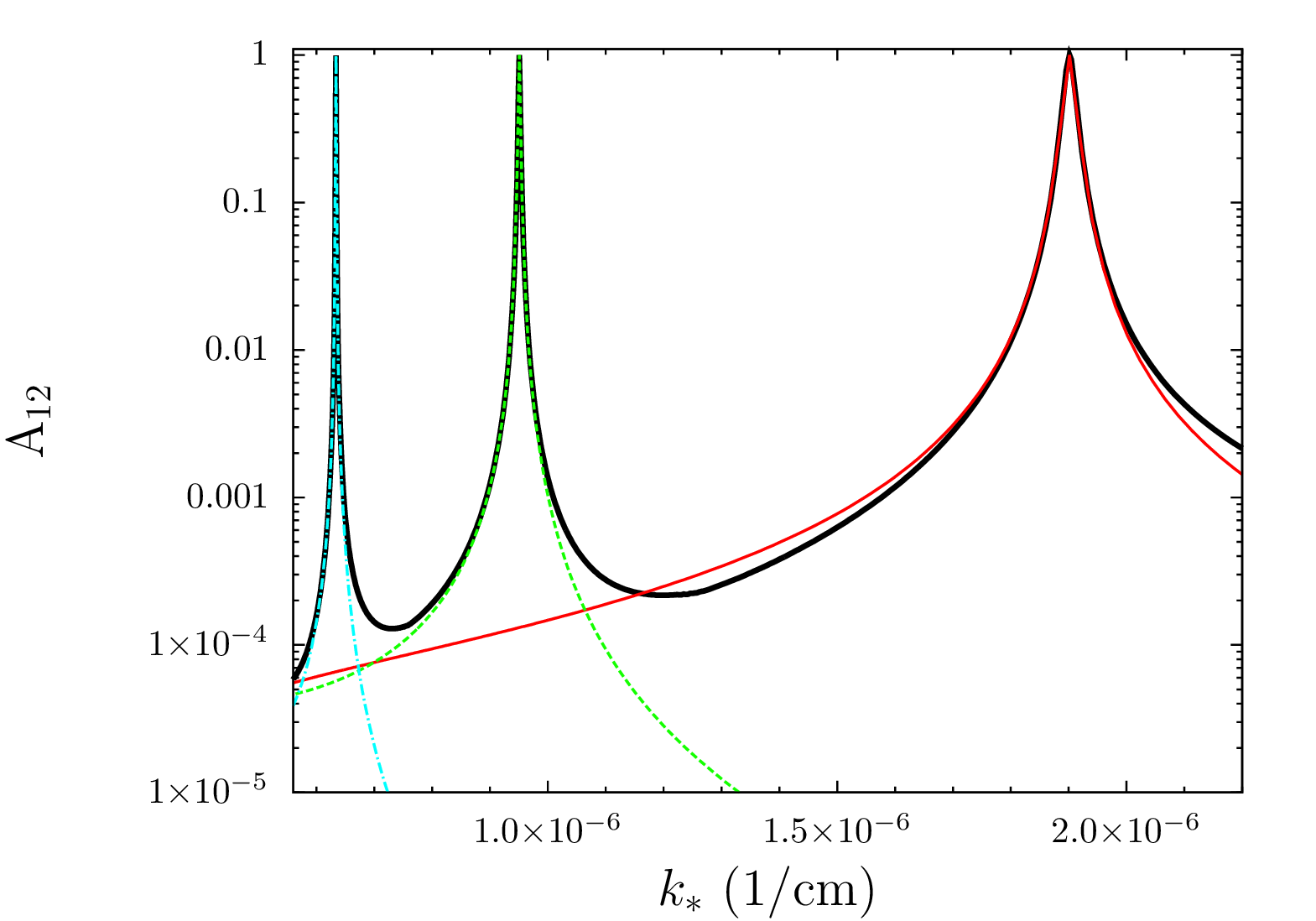}
\caption{A comparison between the Rotating Wave Approximation prediction of the oscillation amplitude for resonances $n\in\{1, 2, 3\}$ ($n=1$ is the rightmost peak) and the amplitude derived from the numerical solution (solid black) as a function of the applied frequency for our test problem.} \label{fig:compare_amplitude}
\end{figure}
To see whether the amplitude and wavelength of the transition are correctly predicted at other values of $k_{\star}$ we scan in this quantity and compare the 
numerically derived results with the analytic RWA predictions for resonance numbers $n\in\{1,2,3\}$. We have changed the amplitude to 
$C_{\star}=0.3$ in order to broaden the resonances but all other parameters remain the same as in the previous test problem.
To determine the wavelength of the numerical results we first Fourier transform the numerically generated $P_{12}(r)$ and define the wavelength to be that at the peak of the longest wavelength Fourier component. The amplitude is found from the numerical results by finding the maximal value of $P_{12}(r)$.  
The results are shown in figures \ref{fig:compare_wavelength} and \ref{fig:compare_amplitude} and again we see that in the vicinity of the parametric resonances 
the analytic and numerical results are in very good agreement. Note the narrowing of the resonances as the resonance number $n$ increases. 
The places where the analytic RWA predictions do not agree with the numerical results occur when the amplitude is very small.  

%
%

\section{Applications}

One can contemplate many situations where neutrino transformation may be stimulated: in a terrestrial experiment, in the Sun, and in supernovae.   
By careful selection of $k_{\star}$ so that $\delta\breve{k}_{ij} + n\,k_{\star}= 0$ for some integer $n$ we maximise the amplitude of the 
transition between the states but at the same time we also minimise the wavenumber $q_{n}=|\kappa_n|$ and maximise the wavelength $\lambda_{n}$. Thus, in order to reduce the length-scale over which the transition occurs we need to maximise $|\kappa_{n}|$ which has the added benefit of increasing the width of the transition resonance. From the expression for
$|\kappa_{n}|$ we can identify three factors to maximise: the product $C_{\star}\,V_{\star}$, the combination $n J_{n}(z_{12})/z_{12}$ and the product
of the matter mixing matrix elements $\breve{U}_{e1}^{\star}\breve{U}_{e2}$. The $n J_{n}(z_{12})/z_{12}$ combination has a maximum value of $1/2$ which occurs
at $z_{12}=0$ for $n=\pm 1$ and at $z_{12} \sim \pm 1$ for all other values of $n$. Likewise the product $\breve{U}_{e1}^{\star}\breve{U}_{e2}$ also has a maximum value of $1/2$
which occurs when $V_{\star}$ is equal to the MSW resonance density for a given neutrino energy. Thus for $n=\pm 1$ it is possible to maximise both terms simultaneously
but for $n\neq \pm 1$ we cannot. Inserting both limits we find $|\kappa_{n}| \leq C_{\star}\,V_{\star}/4$ so if we want to maximise $|\kappa_{n}|$ we need to 
maximise the density and/or the fluctuation amplitude.

If we consider a terrestrial experiment then the typical density of terrestrial material is of order $1\;{\rm g/cm^{3}}$. Using our result for the maximal value of $|\kappa_1|$ 
at the parametric resonance we find $\lambda_{1} = 2\pi/|\kappa_{1}| \gtrsim 10,000\;{\rm km}$ if $C_{\star}=1$. If we instead select a $k_{\star}$ that is off 
resonance by the width $|\kappa_{n}|$ we reduce the transition wavelength by a factor of $2$ but decrease the amplitude by a factor of $4$. An experiment that could observe a stimulated transition 
amplitude of order 1\% would do so over a wavelength of $\sim 1,000\;{\rm km}$. The construction of $\sim 1,000\;{\rm km}$ sinusoidal density profile to observe a 1\% effect makes the difficulty of observing stimulated transitions on Earth extreme. A similar conclusion has been reached by many others \cite{2001PAN....64..787A,2002PhLB..532..259J}  

If we now consider the Sun then we can imagine decomposing the solar profile into an `average' profile and then superposing a Fourier series to account for the departures from the average due to 
convection, turbulence or other features. As a neutrino propagates from the solar core outwards the splitting between its eigenvalues evolves so one might expect that several sinusoidal components might be capable of stimulating neutrino transition during the passage. Ignoring for the present the effect of the evolving underlying profile, adopting a density typical of the solar core i.e.\ $\rho \sim 100\;{\rm g/cm^{3}}$ and density fluctuations of order $C_{\star} \sim 10^{-5}$ gives a transition wavelength of $\lambda_{1} \sim 10^{6}\;{\rm km}$ which is bigger than the solar radius. 

That leaves supernovae. As with the Sun, we can imagine decomposing the profile into an `average' profile and superposing a Fourier series to account for departures from the average. If we focus our considerations upon the region of the H resonance which, for $MeV$ scale neutrino energies, is where the density is of order $\sim 10^{4}\;{\rm g/cm^{3}}$ then fluctuation amplitudes of $C_{\star} \sim 1\%-10\%$ - which are reasonable due to the violent fluid motions found in the explosion - will lead to transition wavelengths which are $\lambda_{1} = 2\pi/|\kappa_{1}| \sim 100-1000\;{\rm km}$, comfortably smaller than the physical size of the star.     
Thus, the combination of large density and large amplitudes means that supernovae appear to be an environment where stimulated transition between matter states can occur. If we are to apply our theory results then
we also have to grapple with the issue of the non-constant underlying profile.   

%
%
\begin{figure}
\includegraphics[clip,width=\linewidth]{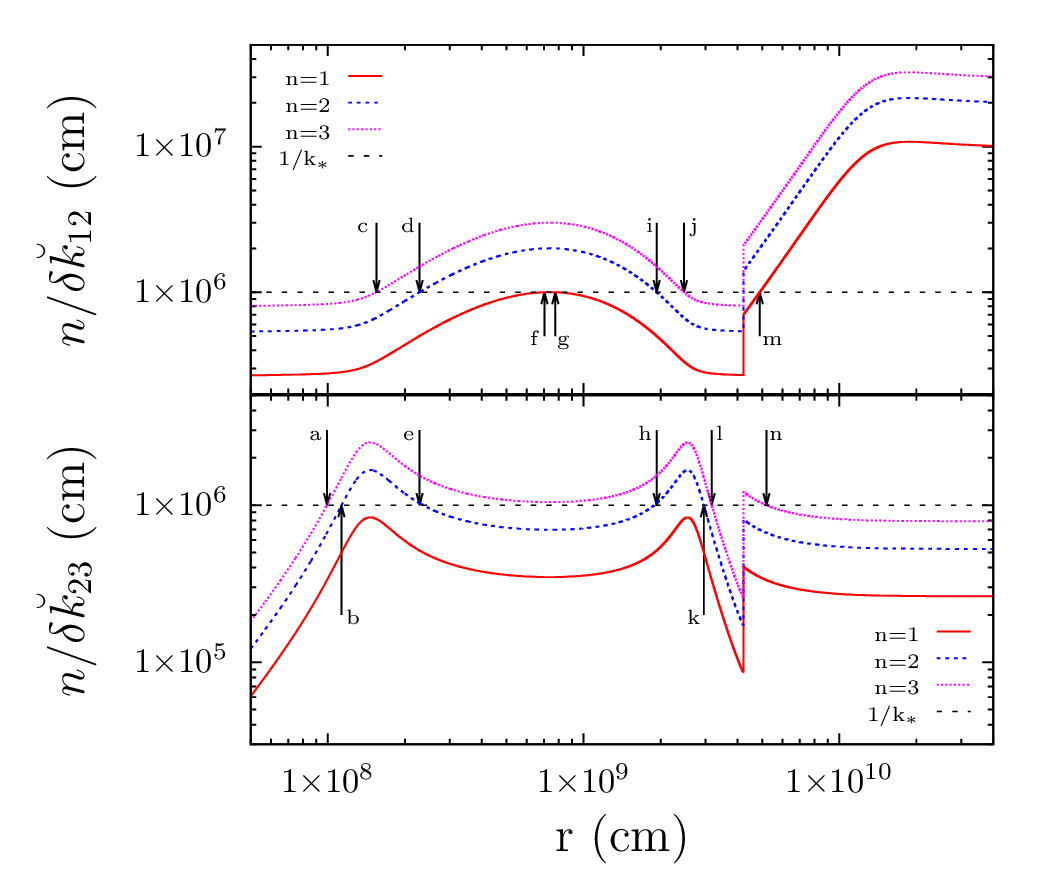}
\caption{Using the base density profile described in the text, we plot multiples of the wavelength 
associated with the mass splittings, $n/\delta \breve{k}_{12}$ (upper panel) and $n/\delta \breve{k}_{23}$ (lower panel) as a function of distance. 
In each panel, the lowest solid curve represents n=1, the middle is n=2 and highest is n=3. The wavelength of a sinusoidal perturbing potential with reduced wavelength $\lambdabar = 1/ k_{\star} = 10\;{\rm km}$ is plotted as the dashed line. When any of the solid lines crosses the dashed line, the parametric resonance condition is fulfilled and thus there exists possibility of a stimulated transition. \label{fig:masssplittings}}
\end{figure}

\subsection{Non-constant Density Profiles: A Supernova Test Problem}
To address the issue of how to apply our results when the underlying profile is not a constant and to explore the possibility that stimulated transitions can occur in supernovae further,   
we construct a density vs. radius relationship from the parametrised form from Fogli \etal \cite{2003PhRvD..68c3005F} with the shock position set at $t=4\;{\rm s}$.
It is not our intention here to embark on a full study of the effect of multiple sinusoids, we shall concentrate on the effect of just one. Thus 
we superpose on the profile a single sinusoid perturbation with a reduced wavelength $\lambdabar=1/k_{\star} = 10\;{\rm km}$ and amplitude $C_{\star}=0.1$. Through this profile we send a neutrino with the following properties, $\delta m_{12}^{2}= 3 \times 10^{-3} {\rm eV^2}$, $\delta m_{23}^{2} = 8 \times 10^{-5} {\rm eV^2}$, $\theta_{12}=33^{\circ}$, $\theta_{13}=9^{\circ}$, $\theta_{23}=45^{\circ}$, $\delta=0$ and with an energy of $20\;{\rm MeV}$. 

\begin{figure}[t]
\includegraphics[clip,width=\linewidth]{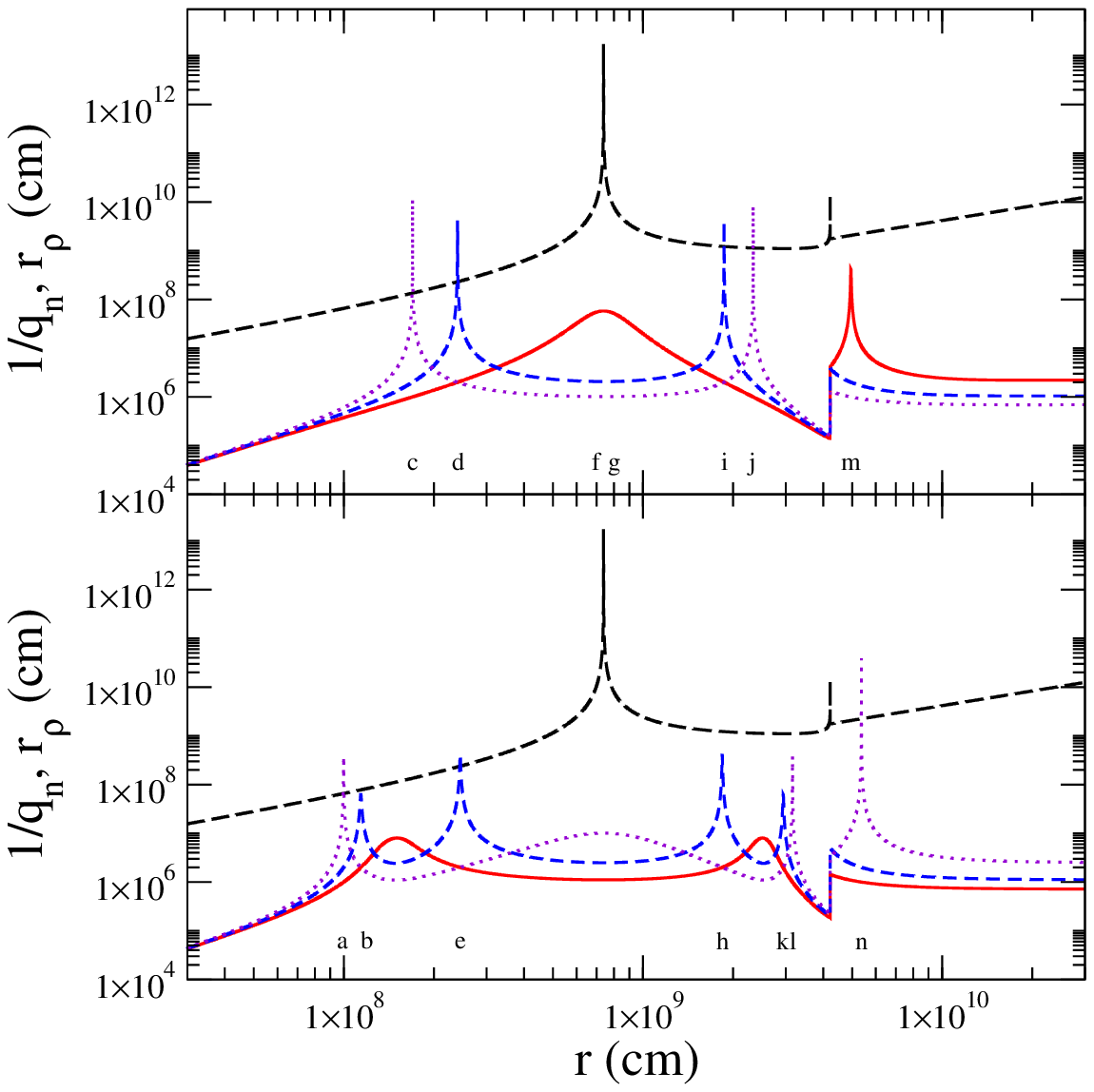}
\caption{Comparison of the reduced wavelength $1/q_n$ for the n=1, n=2 and n=3 modes (red, blue and purple lines) with the density scale height, $r_\rho$ for the base density profile (black dashed line). At the parametric resonances identified in \fref{fig:masssplittings}, the transition wavelength $1/q_n$ becomes long. When  $r_\rho$ is larger than $1/q_n$, we expect stimulated transitions. 
\label{fig:kr_and_scaleheight}}
\end{figure}

Using the base density profile and the parametric resonance condition, $\delta\breve{k}_{ij} + n\,k_{\star} = 0$, where the subscripts $i$ and $j$ denote matter states 1, 2 or 3, we predict the locations where significant stimulated transitions between the states may occur. These are the places where the $ n / \delta\breve{k}_{ij}$ curves in \fref{fig:masssplittings}, intersect the horizontal dashed line corresponding to $\lambdabar$. We see that there are seven locations (c,d,f,g,i,j,m) in the 1,2 channel (top panel) and seven locations (a,b,e,h,k,l,n) in the 2,3 channel (bottom panel).
The formalism developed in the previous section 
assumed a constant density profile, but the supernova-like density profile we have chosen is not constant. Thus we only expect to see stimulated transitions at some subset of the fourteen locations identified in \fref{fig:masssplittings}. To determine this subset, we use a comparison of the density scale height $r_{\rho}=|\rho/ (d\rho/dr)|$ with the reduced parametric resonance wavelength $1/q_n$. When $r_{\rho} \gg 1/q_n$, then the constant density approximation is reasonable, and we should expect to see stimulated transitions. In \fref{fig:kr_and_scaleheight} where the top (bottom) panel again corresponds to the 1,2 (2,3) channel, it can be seen that the stimulated transition wavelength becomes long at the parametric resonance locations. In the 1,2 channel, only for points f,g is $r_{\rho} \gg 1/q_n$, although $r_{\rho}$ is slightly larger than $1/q_n$ at m. Thus, we expect a significant stimulated transition at f,g, and a more moderate one at m. In the 2,3 channel, this figure suggests that the most significant transition will be at location k with more moderate transitions at h and l.
%
%
%
%
\begin{figure}[t]
\includegraphics[clip,width=\linewidth]{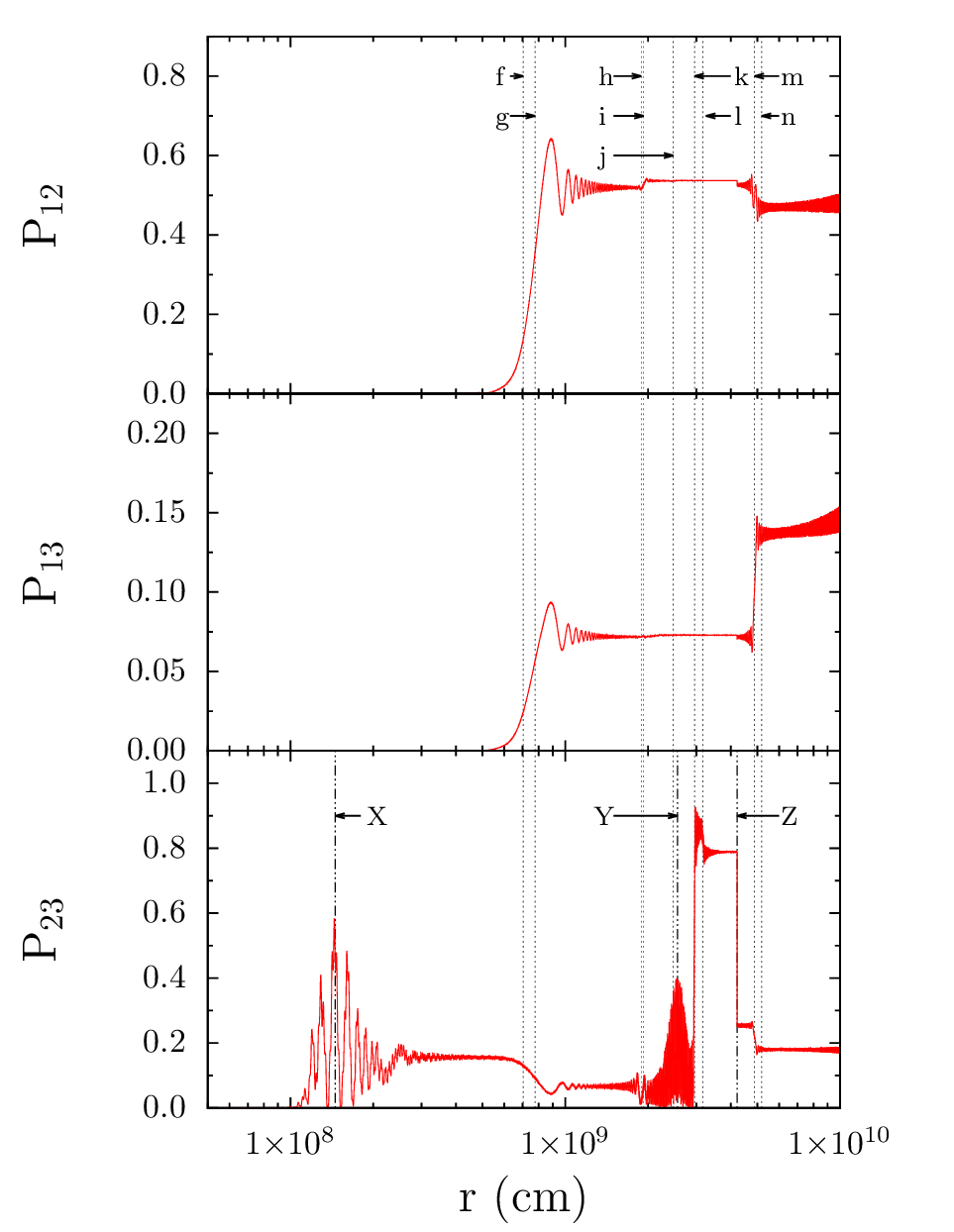}
\caption{Transition probabilities between matter eigenstates, $P_{12}$, $P_{13}$ and $P_{23}$, as calculated numerically. Large stimulated transitions occur at f and g (top panel) and at k (bottom panel). Smaller amplitude stimulated transitions occur at m, h, and l. The transitions labelled X, Y and Z are base profile MSW H resonances. 
\label{fig:probabilities}}
\end{figure}
In \fref{fig:probabilities} we check these results by evolving the neutrinos through the base + perturbing profile with an exact numerical three neutrino calculation. The base density profile possesses three MSW high (H) density resonances. Two of these, labelled X and Y on the figure, are semi-adiabatic while the third, Z, is at the position of the shock and is diabatic. The evolution of the neutrino through the three MSW resonances is accounted for with the unperturbed matrix $\breve{S}$. Note that we have plotted transition probabilities between matter eigenstates not mass eigenstates, so the relevant MSW H transition is in the 2,3 channel.

At the expected stimulated transition locations, we indeed see strong transitions at f,g between matter eigenstates 1 and 2, and a somewhat smaller transition in the same channel at m. In the 2,3 channel we see a strong transition at k and smaller transitions at h and l. We see also a few additional features at the parametric resonance locations caused by simultaneous occurrence of 
stimulated transitions between multiple pairs of states. For example, around points f,g and around m both $1/\delta \breve{k}_{12} \approx \lambdabar$ {\it and} $3/\delta \breve{k}_{23} \approx \lambdabar$. This dual satisfaction of the parametric resonance conditions implies $4/\delta \breve{k}_{13} \approx \lambdabar$ which means we can have stimulated transitions between all three states simultaneously.
Our more complete study of stimulated neutrino transitions due to multiple sinusoids and multiple neutrino flavors is in preparation. 

%
%

\section{Summary and Conclusions}

We have investigated the effect of sinusoidal density fluctuations upon neutrino
propagation and derived an analytic solution for the matter basis transition probabilities in the case of constant density. Large 
transitions between the states can occur and we have derived new expressions for the wavelengths and amplitudes using the Rotating Wave Approximation.
When we examine environments where stimulated transitions may occur we find only supernovae are viable due to the requirement of simultaneous high densities and relatively large amplitudes of perturbing modes.  
To explore this possibility we presented an example calculation with a supernova-like base density profile to which we added a sinusoidal perturbation. When compared with a numerical three neutrino flavor calculation, we found we were able to successfully predict the locations of the stimulated transitions and the strengths despite the presence of multiple MSW resonances in the unperturbed profile. 
A more realistic supernova profile would be more complicated than what we have considered here and it would be necessary to decompose it into a base density profile and a perturbing profile represented as a Fourier series of sinusoidal fluctuations with appropriate coefficients. Since any density profile can be decomposed in this way, one could also consider treating the turbulent aspects of the profile in this fashion. In this latter case fluctuations on many wavelengths will likely have comparable contributions. These cases will be explored in future work.

\ack

This work was supported by DOE grants DE-SC0004786 and DE-SC0006417 (JPK), DE-FG02-02ER41216 (GCM+KMP) and by 
a NC State University GAANN fellowship (KMP). 



\section*{References}

\begin{thebibliography}{99}

\bibitem{Duan:2009cd} Duan H and Kneller J P 2009 J.\ Phys.\ G {\bf 36} 113201 

\bibitem{1996PhRvD..54.6323B} Balantekin A B and Beacom J F 1996 \PR D {\bf 54} 6323 

\bibitem{2001JPhG...27.2405F} Fishbane P M and Kaus P 2001 J.\ Phys.\ G {\bf 27} 2405

\bibitem{2001EPJC...20..507O} Ohlsson T and Snellman H 2001 European Physical Journal C {\bf 20} 507

\bibitem{2004JMP....45.4053B} Blennow M and Ohlsson T 2004 Journal of Mathematical Physics {\bf 45} 4053

\bibitem{Ermilova} Ermilova V K, Tsarev V A and Chechin V A 1986 Kr. Soob, Fiz. Lebedev Institute {\bf 5} 26 

\bibitem{1987PhLB..185..417S} Sch{\"a}fer A and Koonin S E 1987 \PL B {\bf 185} 417 

\bibitem{Akhmedov} Akhmedov E K 1988 Sov. J. of Nuclear Physics {\bf 47} 301, Yad. Fiz {\bf 47} 475  

\bibitem{1989PhLB..226..341K} Krastev P I and Smirnov A Y 1989 \PL B {\bf 226} 341 

\bibitem{PhysRevD.43.2484} Haxton W C and Zhang W-M 1991 \PR D {\bf 43} 2484 

\bibitem{1996PhRvD..54.3941B} Balantekin A B, Fetter J M and Loreti F N 1996 \PR D 54, 3941

\bibitem{2009PhLB..675...69K} Koike M, Ota T, Saito M and Sato J 2009 \PL B {\bf 675} 69  

\bibitem{1999NuPhB.538...25A} Akhmedov E K 1999 Nuclear Physics B {\bf 538} 25  

\bibitem{2001PAN....64..787A} Akhmedov E K 2001 Physics of Atomic Nuclei {\bf 64} 787
 
\bibitem{2002PhLB..532..259J} Jacobsson B Ohlsson T Snellman H and Winter W 2002 Physics Letters B {\bf 532} 259

\bibitem{Loreti:1994ry} Loreti F N and Balantekin A B 1994 \PR D {\bf 50} 4762

\bibitem{Loreti:1995ae} Loreti F N, Qian Y-Z, Fuller G M and Balantekin A B 1995 \PR D {\bf 52} 6664

\bibitem{Kneller:2010ky} Kneller J P Preprint arXiv:1004.1288 [hep-ph]

\bibitem{Kneller:2010sc} Kneller J P and Volpe C 2010 \PR D {\bf 82} 123004 

\bibitem{Friedland:2006ta} Friedland A and Gruzinov A Preprint astro-ph/0607244

\bibitem{2003PhRvD..68c3005F} Fogli G L, Lisi E, Mirizzi A and Montanino D, 2003 \PR D {\bf 68} 033005  

\bibitem{Kneller:2005hf} Kneller J P and McLaughlin G C 2006 \PR D {\bf 73} 056003 

\bibitem{Kneller:2009vd} Kneller J P and McLaughlin G C 2009 \PR D {\bf 80} 053002 

\bibitem{1985PhLA..108..340K} Kmetic M A and Meath W J 1985 \PL A {\bf 108} 340  

\bibitem{Wolfenstein1977} Wolfenstein L 1978 \PR D {\bf 17} 2369 

\bibitem{M&S1986} Mikheev S P and Smirnov A I, (1986) \NC C {\bf 9} 17 

\bibitem{2009PhRvD..80e3002K} Kneller J P and McLaughlin G C 2009 \PR D {\bf 80} 053002
 
\bibitem{2012JPhG...39c5201G} Galais S, Kneller J \& Volpe C 2012 J.\ Phys.\ G {\bf 39} 035201 








 
\end{thebibliography}
\end{document}